\documentclass[sigconf]{acmart}
\usepackage{xspace}
\AtBeginDocument{%
  }

\setcopyright{acmlicensed}

\newcommand{\method}{NEGEL\xspace}

\copyrightyear{2025} \acmYear{2025} \setcopyright{rightsretained} \acmConference[WWW Companion '25]{Companion Proceedings of the ACM Web Conference 2025}{April 28-May 2, 2025}{Sydney, NSW, Australia} \acmBooktitle{Companion Proceedings of the ACM Web Conference 2025 (WWW Companion '25), April 28-May 2, 2025, Sydney, NSW, Australia} \acmDOI{10.1145/3701716.3717806} \acmISBN{979-8-4007-1331-6/2025/04}


\begin{document}

\title{Towards Non-Euclidean Foundation Models: Advancing AI Beyond Euclidean Frameworks}

\author{Menglin Yang}
\affiliation{%
    \institution{Hong Kong University of Science and Technology (GZ)\city{Guangzhou}\country{China}} 
\institution{Yale University\city{New Haven}
\\
\country{United States}}
}
\email{menglin.yang@outlook.com}

\author{Yifei Zhang}
\affiliation{%
    \institution{Nanyang Technological University}
    \country{Singapore}\country{Singapore}
}
\email{yifeiacc@gmail.com}

\author{Jialin Chen}
\affiliation{
\institution{Yale University}
\city{New Haven}
\country{United States}
}
\email{jialin.chen@yale.edu}

\author{Melanie Weber}
\affiliation{
\institution{Harvard University\city{Cambridge}
\\
\country{United States}}
}
\email{mweber@seas.harvard.edu}

\author{Rex Ying}
\affiliation{%
\institution{Yale University\city{New Haven}
\\ \country{United States}}
}
\email{rex.ying@yale.edu}

\renewcommand{\shortauthors}{Menglin Yang, Yifei Zhang, Jialin Chen, Melanie Weber, \& Rex Ying}

\begin{abstract}
In the era of foundation models and Large Language Models (LLMs), Euclidean space is the de facto geometric setting of our machine learning architectures~\cite{vaswani2017attention,he2016deep,touvron2023llama}. 
However, recent literature has demonstrated that this choice comes with fundamental limitations~\cite{linial1995geometry,suzuki2021generalization,bachmann2020constant,kiani2024hardness}. 
To that end, non-Euclidean learning is quickly gaining traction, particularly in web-related applications where complex relationships and structures are prevalent.
Non-Euclidean spaces, such as hyperbolic, spherical, and mixed-curvature spaces, have been shown to provide more efficient and effective representations for data with intrinsic geometric properties, including web-related data like social network topology, query-document relationships, and user-item interactions~\cite{peng2021hyperbolic,mettes2023hyperbolic,yang2022hyperbolic,weber2020robust,ste,zhang2020deep,law2020ultrahyperbolic,xiong2022pseudo,sim2021directed,bombelli1987space,gu2019learning,wang2021mixed,sun2021self,zhu2020graph}. 
Integrating foundation models with non-Euclidean geometries has great potential to enhance their ability to capture and model the underlying structures, leading to better performance in search, recommendations, and content understanding.
This workshop focuses on the intersection of \textbf{N}on-\textbf{E}uclidean Foundation Models and \textbf{GE}ometric \textbf{L}earning (\method), exploring its potential benefits, including the potential benefits for advancing web-related technologies, challenges, and future directions. Workshop page: {\color{blue!50!black}\url{https://hyperboliclearning.github.io/events/www2025workshop}}
\end{abstract}

\maketitle

\section{Workshop Summary}

\textbf{Objectives.} The primary objective of this workshop is to provide a platform for researchers from diverse backgrounds—including \textbf{non-Euclidean representation learning and geometric deep learning, and large foundation models, alongside web-related applications}—to share knowledge, exchange ideas, and discuss the latest advances and challenges in integrating non-Euclidean spaces with these models.

Non-Euclidean (including hyperbolic, spherical, mixed-curvature) representations are particularly valuable for various web-related applications. In \textit{web text representation}, they effectively model hierarchical semantic relationships and latent topic structures that exhibit non-Euclidean properties. In \textit{network analysis}, they capture the complex topologies of web and social graphs, including symmetries, cycles, and community structures. For \textit{recommender systems}, non-Euclidean spaces, especially the hyperbolic space, represent the intricate geometry of user-item interactions. In \textit{information retrieval}, they enable more expressive embeddings for query-document relationships and semantic similarity. 
Additionally, in \textit{web multi-modal representation}, non-Euclidean geometries preserve the inherent structure across different data modalities, improving cross-modal alignment and understanding.

By bridging the gap between foundation models and non-Euclidean machine learning, this workshop aims to unlock new opportunities for advancing AI beyond traditional Euclidean frameworks. 

\textbf{Proposed Duration.} We propose a \textit{full-day workshop}, which will include invited talks, poster sessions, a panel session, a breakout discussion session, and short contributed talks on oral and outstanding paper submissions. The actual workshop schedule will be aligned with the official Web Conference 2025 schedule. 

{\small
\makebox[2cm][r]{8:00--9:00AM}\quad\makebox[0pt][l]{Poster setup}\\
\makebox[2cm][r]{9:00--9:05AM}\quad\makebox[0pt][l]{\textbf{Opening remarks \& Ice-breaking activities}}\\
\makebox[2cm][r]{9:05--10:00AM}\quad\makebox[0pt][l]{\textbf{Invited talk}: Shirui Pan}\\
\makebox[2cm][r]{10:00--11:00AM}\quad\makebox[0pt][l]
{\textbf{Invited talk}: Philip S. Yu}\\
\makebox[2cm][r]{11:00--12:00PM}\quad\makebox[0pt][l]{\textbf{Invited talk}: Smita Krishnaswamy}\\
\makebox[2cm][r]{12:00--1:30PM}\quad\makebox[0pt][l]{Lunch break \& informal discussions}\\
\makebox[2cm][r]{2:00--2:30PM}\quad\makebox[0pt][l]{\textbf{Invited talk}: Pascal Mettes}\\
\makebox[2cm][r]{3:00--3:30PM}\quad\makebox[0pt][l]{\textbf{Invited talk}: Min Zhou}\\
\makebox[2cm][r]{3:30--4:30PM}\quad\makebox[0pt][l]{Interactive coffee break \& breakout discussions}\\
\makebox[2cm][r]{4:30--5:00PM}\quad\makebox[0pt][l]{\textbf{Panel discussions}: FM \& Non-Euclidean Space}\\
\makebox[2cm][r]{5:00--5:30PM}\quad\makebox[0pt][l]{\textbf{Contributed Talks}}
}

\section{Topics and Scope}
The workshop will include submissions, talks, and poster sessions exploring the intersection of foundation models, non-Euclidean machine learning, and graph learning with particular emphasis on web-related applications:

\textbf{(1) Theoretical Foundations}: Recent works have analyzed the generalization error~\cite{suzuki2021generalization}, trainability~\citep{kiani2024hardness}, representation precision~\cite{mishne2023numerical}, and curvature-dimension tradeoff~\cite{alvarado2023curvature,kiani2024hardness2} of non-Euclidean models, yet there remains much to be explored. We welcome contributions that further investigate the theoretical underpinnings of non-Euclidean spaces, examining properties like curvature, geodesics, and isometries, and their effects on the performance of foundation models. 

\textbf{(2) Architectures, Algorithms, and Large Foundation Models}: Preliminary works~\cite{nickel2017poincare,nickel2018learning,ganea2018hyperbolic,hgcn2019,liu2019HGNN,khrulkov2020hyperbolic,tay2018hyperbolic,weber2023global,liu2022enhancing,gu2019learning,wang2021mixed,sun2021self,zhu2020graph,weber2020neighborhood,weber2020robust,kiani2024hardness,Yang2024HypformerEE,sun2025riemanngfm,wang2024mixed,van2023poincare} have established foundational representation models within non-Euclidean spaces. 
We invite submissions that propose novel architectures and algorithms for integrating foundation models with non-Euclidean spaces. 

This includes (i) adapting existing foundation models—such as LLMs, ViTs, and multi-modal models—to work within non-Euclidean geometries or incorporating geometric inductive biases~\cite{Yang2024HyperbolicFF}; (ii) developing new foundation models specifically designed for non-Euclidean spaces; 
(iii) investigating the use of non-Euclidean architectures, such as graph neural networks, hyperbolic neural networks, spherical neural networks, and mixed-curvature models, in combination with foundation models.
(iv) adapting existing foundation models for web tasks in non-Euclidean spaces; 
(v) developing new foundation models specifically for web-centric non-Euclidean representations; and leveraging non-Euclidean architectures for web-specific challenges in search, recommendation, and content understanding.

\textbf{(3) Web Applications and Case Studies}:
Non-Euclidean representations have shown impressive results in graph analysis \cite{bevilacqua2021equivariant,yang2021discrete}, text processing \cite{hgtm}, image understanding \cite{khrulkov2020hyperbolic,van2023poincare}, bioinformatics \cite{hoogeboom2022equivariant}, recommender systems~\cite{yang2022hicf,yang2022hrcf,chen2022modeling} and other areas. We welcome submissions demonstrating applications in web-specific domains, including social network analysis, web knowledge graphs, search systems, recommender systems, and multi-modal web content understanding.

\textbf{(4) Trustworthiness and Robustness}: The development of foundation models and non-Euclidean representations for web applications requires careful consideration of trustworthiness and robustness~\cite{zhang2021we,weber2020robust,suzuki2023tight,liu2024client}. 
We seek submissions addressing adversarial robustness, fairness, interpretability, and privacy in web-related non-Euclidean models, particularly for user-facing applications and sensitive data processing.

\textbf{(5) Benchmarks and Evaluation}:
To advance the development of non-Euclidean foundation models for web applications, we encourage submissions to introduce new web-specific datasets, evaluation protocols, and tools. Particular emphasis will be placed on benchmarks that assess performance on real-world web tasks and scalability to web-related data.

\textbf{Submissions.} We welcome both short research papers of up to 4 pages and full-length research papers of up to 8 pages (excluding references and supplementary materials). All accepted papers will be presented as posters. We will select around 4 papers for oral presentations and 2 papers for outstanding paper awards. 

\textbf{Anticipated audience size.} 
We expect this to be a medium-sized workshop, with 20$\sim$100 attendees, based on our preliminary survey with researchers interested in this topic. There is significant interest in \method from both academia and industry.

\textbf{Details on the poster and discussion session.} Posters will be organized into four thematic groups aligned with core research directions: 
(1) theoretical foundations and algorithms of \method, 
(2) web-relevant models and applications, (3) trustworthiness and robustness , and 
(4) benchmarks and Evaluation. 
Each group will have designated discussion topics centered around challenges and opportunities in \method. We will require 20 poster boards to accommodate all accepted submissions while ensuring comfortable spacing for group discussions. 

During the panel session, we will initiate the discussions with the highlighted questions and transition to the questions from the live audience.

\textbf{Relevance to The Web Conference 2025.}
As web-related data increasingly exhibits complex non-Euclidean structures, from social networks to knowledge graphs, our workshop addresses fundamental challenges in web technologies: enhancing web search through better semantic representations, improving recommendation systems with more expressive user-item relationship modeling, advancing social network analysis with topology-aware embeddings, and enabling more effective multi-modal web content understanding. 

\textbf{How does this workshop differ from previous events?}
A closely related workshop, DiffGeo4DL\footnote{https://sites.google.com/view/diffgeo4dl/}, focused on differential geometry and was held at NeurIPS 2020, five years ago. The recent advances in foundation models motivate revisiting this topic with an emphasis on
web-scale applications of non-Euclidean foundation models. 
While other workshops have focused on geometric deep learning\footnote{e.g., NeurReps (https://www.neurreps.org), Geometrical and Topological Representation Learning (https://gt-rl.github.io/), TAG(https://www.tagds.com/home)}, our workshop uniquely addresses the challenges and opportunities of non-Euclidean representation learning in web technologies, from social network analysis to large-scale recommendation systems and web search. 
This web-centric focus, combined with the integration of foundation models for web applications, provides a novel perspective that distinguishes our workshop in both scope and practical impact.

\section{Workshop Committee}

\subsection{Workshop Organizers}
\textbf{\href{https://scholar.google.com/citations?user=KroqSRUAAAAJ&hl=en}{Menglin Yang (menglin.yang@outlook.com)}}: Menglin Yang currently serves as an Assistant Professor at the Hong Kong University of Science and Technology (Guangzhou). Previously, he conducted postdoctoral research at Yale University. His research focuses on hyperbolic geometric learning and foundation models. He served as the Area Chair of the NeurIPS 2023 GLFrontier workshop and has organized tutorials on hyperbolic representation learning at leading conferences such as KDD 2023 and ECML-PKDD 2022. 

\textbf{\href{https://scholar.google.com/citations?user=DmwXESQAAAAJ&hl=en}{Yifei Zhang (yifei.zhang@ntu.edu.sg)}} Yifei Zhang is a Research Scientist at the College of Computing and Data Science, Nanyang Technological University. His research interests lie broadly in Representation Learning, with a focus on self-supervised, parameter-efficient, and federated approaches using foundation models. 
He is actively involved in academic activities within the field of trustworthy machine learning. Notable contributions include co-delivering a Tutorial on Trustworthy Federated Learning at IJCNN 2023 and serving as a Program Committee member for the Federated Learning at Foundation Models Workshop (FL@FM) at WWW 2024. 

\textbf{\href{https://cather-chen.github.io/}{Jialin Chen (jialin.chen@yale.edu)}}: Jialin Chen is a third-year Ph.D. student at Yale University.
Including her as an organizer not only enriches the diversity of our team (including speakers) in terms of gender and academic representation (including Ph.D. students, postdocs, assistant professors, and professors) but also ensures that the workshop remains inclusive and accessible to early-career researchers. 
Besides, she has demonstrated strong organizational skills in previous workshops and conferences, such as LoG 2022 and 2023. 

\textbf{\href{http://melanie-weber.com/}{Melanie Weber (mweber@seas.harvard.edu)}}: Melanie Weber is an Assistant Professor of Applied Mathematics and of Computer Science at Harvard University. Her research utilizes geometric structures in data to design efficient machine learning methods, focusing on representation learning, machine learning in non-Euclidean spaces, optimization on manifolds, and discrete curvature applications in graph machine learning. 
Her expertise in non-Euclidean space and her extensive academic background will significantly enhance the proposed workshop's depth and reach. 

\textbf{\href{https://scholar.google.com.au/citations?user=6fqNXooAAAAJ&hl=en}{Rex Ying (rex.ying@yale.edu)}}: Rex Ying is an Assistant Professor at Yale University. His research encompasses algorithms for graph neural networks, geometric embeddings, and explainable models. He is known for developing widely used GNN algorithms such as Hyperbolic GNN, GraphSAGE, PinSAGE, and GNNExplainer. 
He has been a committee member of AAAI, ICML, NeurIPS, KDD, WWW for five years and served as area chair for LoG 2022. Rex has led and co-organized several workshops, including the Simulation for Deep Learning Workshop at ICLR 2021, the Graph Representation Learning Workshop at ICML 2020, and the New Frontiers in Graph Learning Workshop (GLFrontiers) at NeurIPS 2022 and 2023. 

\subsection{Invited speakers} 
The invited speakers listed below are scheduled to deliver their talks. Due to the page limitation, their biographies are introduced briefly. For detailed information, please refer to their homepage.

\textbf{\href{https://cs.uic.edu/profiles/philip-yu/}{Philip S. Yu}.} Philip S. Yu is a Distinguished Professor in the Department of Computer Science at UIC and also holds the Wexler Chair in Information and Technology. 
His main research interests include big data, data mining (especially graph/network mining), social networks, privacy-preserving data publishing, data stream, database systems, and Internet applications and technologies.
Dr. Yu has published more than 970 papers in refereed journals and conferences with more than 74,500 citations and an H-index of 127. 

\textbf{\href{https://shiruipan.github.io/}{Shirui Pan}.} Shirui Pan is a Professor and an ARC Future Fellow with the School of Information and Communication Technology, Griffith University, Australia. He has made contributions to advanced graph machine learning methods for solving hard AI problems for real-life applications, including graph classification, anomaly detection, recommender systems, and multivariate time series forecasting.

\textbf{\href{https://scholar.google.com/citations?user=sMQxA3AAAAAJ&hl=de}{Pascal Mettes}.}  Pascal Mettes is an Assistant Professor at the University of Amsterdam working on hyperbolic and hierarchical learning for computer vision.
He organized the ECCV 2022 and CVPR 2023 tutorials on hyperbolic deep learning and the ICCV 2021 workshop on structured video representation learning. 
Research examples include hyperbolic embeddings for image segmentation, fully hyperbolic ResNets, hyperbolic random forests, and the first open-source library for easy-to-use hyperbolic learning.

\textbf{\href{https://scholar.google.com/citations?user=P8WYyYIAAAAJ}{Min Zhou}.} Min Zhou is currently a Principal Researcher at Huawei, Shenzhen, China.
She has published multiple works related to hyperbolic graph representation learning and hyperbolic recommender systems at top conferences. She delivered tutorials on Hyperbolic Graph Representation Learning at KDD 2023 and ECML-PKDD 2022.

\textbf{Smita Krishnaswamy.} Smita Krishnaswamy is an Associate professor in Genetics and Computer Science at Yale University. She is affiliated with the applied math program, computational biology program, Yale Center for Biomedical Data Science and Yale Cancer Center. Her lab works on the development of machine learning techniques to analyze high dimensional high throughput biomedical data. Her focus is on unsupervised machine learning methods, specifically manifold learning and deep learning techniques for detecting structure and patterns in data.

\subsection{Program Committee}
We are pleased to present the list of researchers who have agreed to serve as Program Committee for our proposed workshop. Their expertise and dedication will ensure a rigorous and insightful review process:
Smita Krishnaswamy (Yale University),
Irwin King (CUHK),
Tinglin Huang (Yale University),
Harshit Verma (Birla Institute of Technology and Science),
Buze Zhang (XJTU), 
Jindong Li (HKUST(GZ)),
and Jiahong Liu (CUHK).
We will broaden our invitation to more program committee members based on submissions, ensuring each submission receives at least two reviews.

\section{Diversity and Inclusion Statement}
\textbf{Diversity of topics.} Our workshop is highly interdisciplinary by design. \method spans multiple core areas of web technologies and their intersecting fields, including information retrieval, web mining, social network analysis, and recommender systems. The approach also bridges fundamental research areas such as machine learning, natural language processing, computer vision, and network science, particularly in their applications to web-related challenges. 
\textbf{Diversity of organization committee.} The organizing committee consists of researchers with a wide variety of demographic backgrounds, and we aim to promote diversity along several axes, including affiliation, gender, seniority, experience, and research areas. 
Among the 5 organizers, there are 1 postdoctoral researchers, 3 assistant professors, and 1 Ph.D. student. Organizers are affiliated with different institutions, including Yale\&HKUST(GZ), Harvard, and NTU. Two of them identify themselves as female and 3 of them as male. 
\textbf{Diversity of speakers.} We similarly promote diversity in the invited speakers. The invited speakers are affiliated with diverse institutions, including the University of Amsterdam, Huawei Noah's Ark Lab, Yale University, Griffith University, and the University of Illinois at Chicago, etc, holding diverse titles including assistant professor, principal researcher, full professor, distinguished professor, etc. Furthermore, 2 of our invited speakers identifies herself as female, and 3 as male. 
\textbf{Diversity of participants.} Our workshop is devoted to providing a welcoming atmosphere for participants with diverse backgrounds. We plan to create a mentorship program, where newcomers to \method can get connected with experienced researchers and learn from each other. 
\bibliographystyle{ACM-Reference-Format}
\bibliography{sample-base}
\end{document}